\documentclass[prb,twocolumn,superscriptaddress,preprintnumbers,a4paper,amsmath,amssymb,showpacs,floatfix]{revtex4}
\usepackage{graphicx}
\usepackage{color}
\usepackage{bm}
\usepackage{epsfig}

\begin{document}

\newcommand{\bb}{$\mathbf{b}^{*}$}
\newcommand{\TNTb}{$T_{\rm N}^{\rm Tb}$}
\newcommand{\etal}{\textit{et al.}}
\newcommand{\pa}{$\mathbf{P}\|a$}
\newcommand{\pc}{$\mathbf{P}\|c$}
\newcommand{\Ha}{$\mathbf{H}\|a$}
\newcommand{\Hb}{$\mathbf{H}\|b$}
\newcommand{\Hc}{$\mathbf{H}\|c$}
\newcommand{\Ps}{$P_{s}$}
\newcommand{\btau}{\mbox{\boldmath$\tau$}}

\title{Ga substitution as an effective variation of Mn-Tb coupling in multiferroic TbMnO$_{3}$}

\author{O.~Prokhnenko}
\affiliation{Helmholtz-Zentrum-Berlin f\"{u}r Materialien und Energie, Glienicker Str.~100, D-14109 Berlin, Germany}

\author{N.~Aliouane}
\affiliation{Helmholtz-Zentrum-Berlin f\"{u}r Materialien und Energie, Glienicker Str.~100, D-14109 Berlin, Germany}
\affiliation{Physics Department, Institute for Energy Technology, P.O. Box 40, NO-2027 Kjeller, Norway}

\author{R.~Feyerherm}
\affiliation{Helmholtz-Zentrum-Berlin f\"{u}r Materialien und Energie, BESSY, D-12489 Berlin, Germany}

\author{E.~Dudzik}
\affiliation{Helmholtz-Zentrum-Berlin f\"{u}r Materialien und Energie, BESSY, D-12489 Berlin, Germany}

\author{A.U.B.~Wolter}
\affiliation{Helmholtz-Zentrum-Berlin f\"{u}r Materialien und Energie, BESSY, D-12489 Berlin, Germany}

\author{A.~Maljuk}
\affiliation{Helmholtz-Zentrum-Berlin f\"{u}r Materialien und Energie, Glienicker Str.~100, D-14109 Berlin, Germany}

\author{K.~Kiefer}
\affiliation{Helmholtz-Zentrum-Berlin f\"{u}r Materialien und Energie, Glienicker Str.~100, D-14109 Berlin, Germany}

\author{D.~N.~Argyriou}
\affiliation{Helmholtz-Zentrum-Berlin f\"{u}r Materialien und Energie, Glienicker Str.~100, D-14109 Berlin, Germany}

\date{\today}


\begin{abstract}
Ga for Mn substitution in multiferroic TbMnO$_{3}$ has been performed in order to study the influence of Mn-magnetic ordering on the Tb-magnetic sublattice. Complete characterization of TbMn$_{1-x}$Ga$_x$O$_{3}$ ($x$ = 0, 0.04, 0.1) samples, including magnetization, impedance spectroscopy, and x-ray resonant scattering and neutron diffraction on powder and single crystals has been carried out. We found that keeping the same crystal structure for all compositions, Ga for Mn substitution leads to the linear decrease of $T_{\rm N}^{\rm Mn}$ and $\tau^{\rm Mn}$, reflecting the reduction of the exchange interactions strength $J_{\rm Mn-Mn}$ and the change of the Mn-O-Mn bond angles. At the same time, a strong suppression of both the induced and the separate Tb-magnetic ordering has been observed. This behavior unambiguously prove that the exchange fields $J_{\rm Mn-Tb}$ have a strong influence on the Tb-magnetic ordering in the full temperature range below $T_{\rm N}^{\rm Mn}$ and actually stabilize the Tb-magnetic ground state.
\end{abstract}

\maketitle

\section{Introduction}

Rare earth manganites $R$MnO$_{3}$ ($R$ = Gd, Tb, Dy, Ho) with orthorhombically distorted perovskite structure have been attracting a lot of attention due to a variety of unusual physical properties including a potentially interesting cross coupling of magnetism and ferroelectricity (FE).\cite{most,kimura,goto,lorenz} These so called multiferroics are of particular interest for understanding the fundamental physical links between spin, charge and lattice degrees of freedom that give rise to magnetoelectric coupling, as well as because of the promising possibility of using these coupled order parameters in novel device applications by controlling the material's polarization state with either electric or magnetic field.\cite{kimura3,fiebig:review,yama}

While it has been clearly shown, that the ferroelectricity in these compounds is magnetically driven, the role of  magnetic 3$d$ and 4$f$ ions was found to be rather different. As commonly accepted, the origin of multiferroicity in $R$MnO$_{3}$ is a result of a cycloidal Mn-magnetic ordering with inverse Dzyaloshinsky-Moriya interaction being the driving force of polar lattice distortions.\cite{kenz,katsura:057205,mostovoy:067601,dagotto,xiang,malash} Spontaneous electric polarization can be therefore expressed via the $m_{y}$ and $m_{z}$ components of the Mn-spins of the cycloid and the magnetic propagation vector, {\btau}, as $\mathbf{P}_{s}\propto m_{y}m_{z}(\mathbf{e_x}\times\btau)$, where $\mathbf{e_x}$ is the unit vector along the axis of rotation of Mn-spins.\cite{mostovoy:067601} Magnetic rare earths  in turn, may determine the polarization direction via their interaction with Mn-spins. In compounds with rare earths  showing strong magnetic anisotropy, the polarization is oriented along $c$-axis, \pc~for $R=$ Tb and Dy, while in those with nonmagnetic (Eu,Y) or less anisotropic Gd, \pa.\cite{hemb,kimura3} In some cases, they even contribute to the magnitude of the polarization as has been shown for DyMnO$_{3}$.\cite{goto,prok} A necessary condition for this is the ordering of rare earths with the same propagation vector as Mn, behavior confirmed now for $R$MnO$_{3}$ with $R$ = Gd, Tb, Dy and Ho above their own ordering temperatures, $T_{\rm N}^{\rm R} < 10$ K.\cite{kenz,feyerherm,munoz} Here, Mn-magnetic sublattice polarizes $R$-spins and the induced $R$-moment is defined by the strength of exchange interaction between Mn- and $R$-ions, $J_{\rm Mn-R}$.

As proposed in Ref.\cite{prok2} the ordering of the $R$-spins below $T_{\rm N}^{\rm R}$ (often referred as $individual$) is strongly dependent on the relative strengths of exchange interactions between Mn- and $R$-ions, $J_{\rm Mn-R}$, and between rare earths themselves, $J_{\rm R-R}$. 
will not significantly affect the $R$-ordering as found for $R$ = Dy that orders on its own ($\tau^{\rm Dy}=\frac{1}{2}\neq\tau^{\rm Mn}$) below 6 K.\cite{feyerherm,prok} On the other hand, strong $J_{\rm Mn-R}$ 
will force the $R$-ordering to the same periodicity as Mn down to the lowest temperatures as in case of $R=$ Ho ($\tau^{\rm Ho} = \tau^{\rm Mn}$).\cite{munoz,brinks}

The most interesting case, however, is an intermediate coupling regime as observed in TbMnO$_{3}$. Here, the Tb- and Mn- orderings remain coupled down to the lowest temperatures through the harmonic coupling of their wave vectors, $3 \tau^{\rm Tb} - \tau^{\rm Mn} = 1$. This is a result of adjustment of Ising-like Tb spins to periodic Mn ordering minimizing system's energy and leading to rather complex Tb-spin structure at low temperatures.\cite{prok2} This intermediate coupling regime should be very sensitive to any variation of $J_{\rm Mn-R}$ and, correspondingly, strong effects on the Tb-magnetic ordering are expected.\cite{prok2}

In this paper we report on combined neutron diffraction, x-ray resonant magnetic scattering, and single crystal magnetization and dielectric measurements on substituted TbMn$_{1-x}$Ga$_{x}$O$_{3}$ ($x$ = 0, 0.04, 0.1) compounds. Substitution of Mn$^{3+}$ ion by the closest isovalent but nonmagnetic Ga$^{3+}$ was chosen to provide minimal disturbance of the lattice and effectively  vary the total $J_{\rm Mn-Mn}$ and, consequently, $J_{\rm Mn-R}$, while keeping $J_{\rm R-R}$ constant. We show that, while keeping the same crystal structure for all compositions, Ga for Mn substitution leads to the linear decrease of $T_{\rm N}^{\rm Mn}$ and $\tau^{\rm Mn}$ reflecting the intended decrease of $J_{\rm Mn-Mn}$ and the increase of the Mn-O-Mn bond angles. At the same time we observe a strong suppression (for $x = 0.04$) and disappearance (for $x \geq 0.1$) of both the induced and the individual Tb-magnetic ordering. This behavior confirms that the exchange fields $J_{\rm Mn-Tb}$ from the Mn have a strong influence on the Tb-magnetic ordering in the full temperature range below the ferroelectric transition and actually stabilize the Tb-magnetic ground state.

\section{Experiment}

Polycrystalline TbMn$_{1-x}$Ga$_{x}$O$_{3}$ ($x$ = 0, 0.04, 0.1) samples were prepared from a mixture of Tb$_4$O$_7$ (4N), Mn$_2$O$_3$ (3N) and Ga$_2$O$_3$ (4N) using standard solid state reaction. The lowest content of Ga ($x\backsimeq$ 4\%) was chosen such as to obtain a homogeneous distribution of Ga within the sample.  The highest content ($x\backsimeq$ 10\%) was determined by our ability to obtain a single phase sample at this composition without significant changes in its crystal structure and magnetic properties. The single crystals were grown in argon atmosphere by the floating zone technique in the IR-heated image furnace (NEC) equipped with two halogen lamps (500W) and double ellipsoidal mirrors.  The growth and the feed rates were maintained to obtain a stable molten zone.

Crystal structure and magnetic ordering of TbMn$_{1-x}$Ga$_{x}$O$_{3}$ were investigated between 2-300~K by neutron diffraction (ND) on powder and single crystal samples using the high resolution E9 diffractometer ($\lambda = $1.797~\AA\ and 2.816~\AA) and the E1 thermal triple axis spectrometer operated in two-axis mode ($\lambda = $2.44~\AA), respectively, at the Helmholtz-Zentrum-Berlin (HZB). The data were analyzed with the FullProf refinement package.\cite{carvajal} Single crystal x-ray magnetic resonant scattering (XRMS) on TbMn$_{1-x}$Ga$_{x}$O$_{3}$ were conducted at the 7~T multipole wiggler beamline MAGS, also HZB.\cite{feyerherm,dudzik}

Physical characterization of the samples was performed on both single crystal and polycrystalline samples. The single crystals were aligned along their principal direction by x-ray Laue diffraction and cut in orthorhombic shapes with each face normal (within 2$^{\circ}$) to a principal crystallographic direction. 

Temperature dependence of low and high field magnetization along three main crystallographic directions were measured on TbMn$_{1-x}$Ga$_{x}$O$_{3}$ ($x$ = 0, 0.04, 0.1) in the VSM mode of a Physical Properties Measurement System (Quantum Design PPMS). The dielectric measurements were performed in a 14.5 T Oxford Instruments cryomagnet equipped with a $^3$He Heliox insert. Opposite faces of the plate-like samples were sputtered with gold  and located between two gold covered copper plates. The top plate and the sample were pressed to the fixed and thermalized bottom plate by a soft spring. The capacitance of such a plate capacitor with a sample as a dielectric was measured by an Andeen-Hagerling 2500 A capacitance bridge working at a frequency of 1 and 20 kHz.

\section{Results and discussion}

\subsection{Crystal structure}

\begin{table*}
\begin{center}
\begin{ruledtabular}
 \begin{tabular}{ccccccc}
  Temperature  &  \multicolumn{3}{c}{300  K}& \multicolumn{3}{c}{12 K}\\ \hline
    $x$           &       0       &      0.04      &    0.10      &          0      &     0.04       &    0.10       \\ \hline
    a (\AA)      &   5.30159(4)  &  5.30049(7)    &   5.29981(7) &     5.31606(4) &  5.31406(6)    &   5.31268(7) \\
    b (\AA)      &   5.85557(4)  &  5.83318(7)    &   5.81546(7) &     5.82304(5) &  5.80384(8)    &   5.78569(7) \\
    c (\AA)      &   7.40003(6)  &  7.41209(9)    &   7.42329(9)  &    7.38443(7)  &  7.39939(8)    &  7.41051(11)  \\ \hline
    x(Tb)          &    -0.0150(3) &    -0.0160(4)  &    -0.0162(4)&     -0.0162(4) &    -0.0159(4)  &    -0.017(4)\\
    y(Tb)          &     0.0814(3) &     0.0801(3)  &     0.0796(4)&      0.0789(3) &     0.0790(3)  &     0.0778(4)\\
    x(O$_{1}$)     &     0.1055(4) &     0.1064(5)  &     0.1050(5)&      0.1064(4) &     0.1058(4)  &     0.1059(5)\\
    y(O$_{1}$)     &     0.4658(4) &     0.4673(4)  &     0.4667(5)&      0.4666(4) &     0.4686(4)  &     0.4678(4)\\
    x(O$_{2}$)     &     0.7035(3) &     0.7038(4)  &     0.7029(4)&      0.7037(3) &     0.7038(3) &     0.7046(4)\\
    y(O$_{2}$)     &     0.3276(3) &     0.3267(3)  &     0.3251(3)&      0.3265(3) &     0.3247(3)  &     0.3235(4)\\
    z(O$_{2}$)     &     0.0515(2) &     0.0514(2)  &     0.0515(2)&      0.0515(2) &     0.0519(2)  &     0.0519(2)\\
    Occ(Ga)        &       0       &     0.044(2)   &     0.104(6) &        0       &     0.044      &     0.104 \\ \hline
$\tau^{\rm Mn}$ (\bb) &        &                &              & 0.2754(2)  & 0.2651(3)  & 0.2463(5) \\
$m_{\rm x}^{\rm Mn}$        &          &                &              &            &           &           \\
$m_{\rm y}^{\rm Mn}$        &          &                &              &  3.95(4)    &   3.30(4) &  3.19(8)  \\
$m_{\rm z}^{\rm Mn}$        &          &                &              &   2.46(10)  &   1.86(15) &  0(1)   \\ \
$\tau^{\rm Tb}$ (\bb) / 2 K &          &                &              &     3/7     &                &       \\
$m_{\rm x}^{\rm Tb}$  / 2 K &          &                &              &   5.79(6)   &           &           \\
$m_{\rm y}^{\rm Tb}$  / 2 K &          &                &              &   3.11(8)   &           &    \\
$m_{\rm z}^{\rm Tb}$  / 2 K &          &                &              &       0     &           &         \\ \hline  
R$_{\rm nucl}$              &  4.67    &     3.83       &     3.79     &      4.14   &    3.26   &    5.47   \\ 
R$_{\rm mag}^{\rm Mn}$      &          &                &              &    6.99     &    6.47   &    7.0    \\
R$_{\rm mag}^{\rm Tb}$      &          &                &              &    13.7      &           &          \\
$\chi^{2}$                  &  1.97    &     3.17       &     2.62     &      1.69   &    3.49   &     3.33   \\
 \end{tabular}
\caption{Crystal and magnetic structure parameters for TbMn$_{1-x}$Ga$_{x}$O$_{3}$ ($x$ = 0, 0.04, 0.1) obtained from the Rietveld refinements of powder ND at room and low temperatures using 1.797~\AA\ and 2.816~\AA. The atom positions are given for $Pbnm$ settings where Mn occupies $4b$ (1/2 0 0), Tb and O$_{1}$ $4c$ ($x$ $y$ 1/4), and O$_{2}$ $8d$ ($x$ $y$ $z$) Wyckoff positions. The low temperature data for all samples are given for 12 K except of Tb-magnetic structure parameters given for 2 K. It is caused by difficulties in fitting the Mn-magnetic reflections being overlapped with Tb-magnetic peaks below 8 K.} \label{table1}
\end{ruledtabular}
\end{center}
\end{table*}

\begin{figure}[bt!]
\begin{center}
\includegraphics[width=85mm]{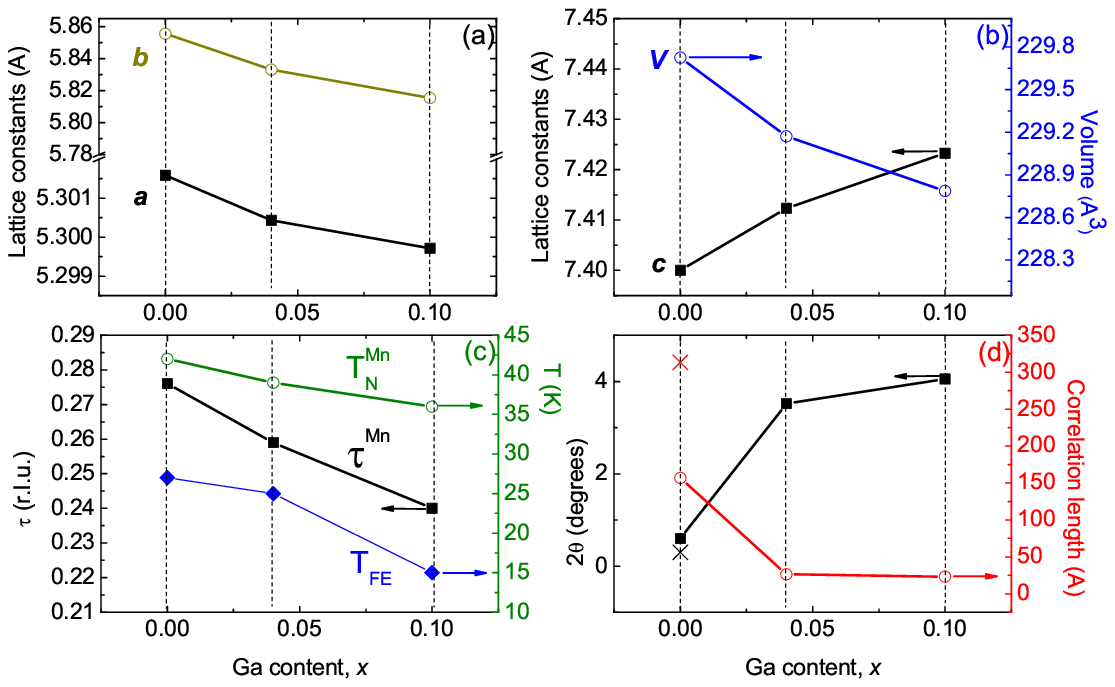}
\caption{(Color online) Magneto-structural parameters of TbMn$_{1-x}$Ga$_{x}$O$_{3}$ compounds as determined from ND:  lattice parameters and volume (a)-(b) at RT, magnetic propagation vector at 2 K, Mn N\'{e}el temperature and ferroelectric transition temperature (c), FWHM of Tb magnetic reflections and correlation length of Tb clusters at 2 K (d) as a function of refined Ga-content. Coherence length has been estimated using the Scherrer formula. Cross symbols in panel (d) correspond to instrument resolution at the position of the strongest Tb-magnetic reflection} \label{fig1}
\end{center}
\end{figure}

Starting with the structural analyses, powder ND measurements at room temperature show that all TbMn$_{1-x}$Ga$_{x}$O$_{3}$ samples are single phase 
and crystallize with the orthorhombically distorted perovskite structure (space group P$bnm$). Ga shares 4$b$  atomic position (Wyckoff notation) with Mn and its refined content is in good agreement with the nominal one (Table~\ref{table1}). We would like to emphasize that despite to close scattering lengths of Tb (7.38 fm) and Ga (7.29 fm), sharp difference between those of Mn (-3.73 fm) and Ga allows to determine their occupancies confirming that all Ga is accounted in the Mn-site.
With increasing Ga-content up to 10\% (Fig.~\ref{fig1}a-b), the lattice linearly contracts along $b$-direction, $\Delta b/b \sim 0.7\%$, and expands along the $c$-axis, $\Delta c/c \sim 0.3\%$ while leaving the $a$-parameter almost constant, $\Delta a/a < 0.1\%$.  The resultant decrease in volume is about 0.4$\%$, approaching the volume of heavier $R$ - DyMnO$_{3}$. Anisotropic change of the lattice, however, leads to an increase of Mn-O-Mn bond angle (by $\sim 0.4^\circ$) as if moving the system to the lighter $R$ - GdMnO$_{3}$. Disturbance of the original TbMnO$_{3}$ lattice is also reflected in changes of Mn-O and Tb-O interatomic distances, both having maximal $\Delta d/d \sim 1.1\%$ as presented in Table~\ref{table2}.

\begin{table}
\begin{center}
\begin{ruledtabular}
 \begin{tabular}{cccc}
    $x$           &       0       &      0.04      &    0.10               \\ \hline
  Mn-O$_{1}$ (\AA) &     1.9431(7) &     1.9463(8)  &    1.9471(8)         \\ 
  Mn-O$_{2}$ (\AA) &    1.9069(16) &     1.9060(20)  &    1.9131(20)       \\ 
  Mn-O$_{2}$ (\AA) & 2.2331(17) &    2.2235(18)  &    2.2084(18)      \\ \hline                          
$\widehat{MnO_{2}Mn}$ ($^{\circ}$) & 144.99(7) &  145.15(8)  & 145.22(8)   \\ 
$\widehat{MnO_{1}Mn}$ ($^{\circ}$) & 144.39(3) & 144.38(3)   &   144.77(3)    \\ \hline                               
  Tb-O$_{1}$ (\AA) &      3.661(3) &      3.633(3)  &      3.622(4)        \\ 
  Tb-O$_{1}$ (\AA) &      2.340(3) &      2.350(3)  &      2.341(4)        \\ 
  Tb-O$_{1}$ (\AA) &      3.203(3) &      3.198(3)  &      3.189(3)       \\ 
  Tb-O$_{1}$ (\AA) &      2.274(3) &      2.269(3)  &      2.276(3)       \\ 
  Tb-O$_{2}$ (\AA) &      2.542(2) &      2.538(2)  &      2.535(3)      \\ 
  Tb-O$_{2}$ (\AA) &      2.570(2) &      2.578(2)  &      2.582(2)       \\ 
  Tb-O$_{2}$ (\AA) &      2.317(2) &      2.311(2)  &      2.311(3)      \\ 
  Tb-O$_{2}$ (\AA) &      3.666(2) &      3.655(2)  &      3.648(3)     \\ 

 \end{tabular}
\caption{Selected distances and angles from Rietveld refinements of powder ND data for TbMn$_{1-x}$Ga$_{x}$O$_{3}$ ($x$ = 0, 0.04, 0.1) at RT.} 
\label{table2}
\end{ruledtabular}
\end{center}
\end{table}

\subsection{Magnetic and dielectric properties}

\begin{figure}
\begin{center}
\includegraphics[width=90mm]{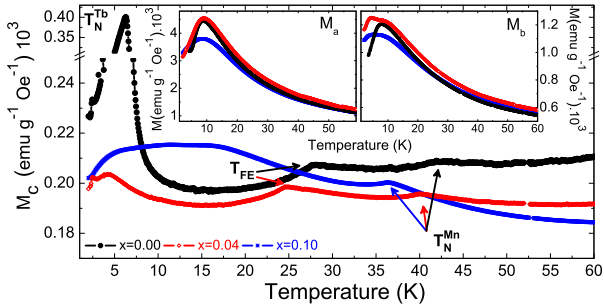}
\caption{a) Temperature dependence of magnetic d.c. susceptibility for TbMn$_{1-x}$Ga$_{x}$O$_{3}$ ($x=$ 0, 0.04 and  0.1) single crystals measured at 1 kOe along $c$-direction. The insert represents the d.c. susceptibility of TbMn$_{1-x}$Ga$_{x}$O$_{3}$ single crystals measured along the $a$- and $b$-crystallographic direction with $H=$ 5 kOe. Data have been collected on heating after previous zero field cooling down to T=2 K.} \label{MvsT-data}
\end{center}
\end{figure}

Turning to the magnetic properties, in Fig.~\ref{MvsT-data} we present the temperature dependence of the magnetization of TbMn$_{1-x}$Ga$_{x}$O$_{3}$ single crystals measured along the $a$- and $b$-crystallographic directions with $H$ = 5~kOe and along $c$ with 1~kOe. We start with magnetization curves measured with $H\| c$ as they show anomalies corresponding to the onset of both Mn- and Tb-magnetic orders. As previously reported for the $x =0$ compound, $M_{c}$ presents three sharp anomalies at $T_{\rm N}^{\rm Mn} = 42$~K, $T_{\rm FE} = 27$~K and \TNTb  $ = 7$~K attributed to the Mn N\'{e}el transition temperature, the transition temperature to the Mn-cycloidal state associated with development of ferroelectricity  
and the Tb N\'{e}el transition temperature, respectively.\cite{kimura} The 4$\%$ Ga-containing compound preserves all three transitions. The first two are shifted down to 39~K and 25~K whereas the last anomaly is enormously reduced and has a maximum around 4.5~K. The 10\%-Ga sample shows only one sharp transition at $T_{\rm N}^{\rm Mn}=36$~K and broad anomaly at around 15 K which might reflect the remanent ferroelectric transition $T_{\rm FE}$. Any signature of \TNTb ~is clearly absent at this composition. 

Magnetization curves along $a$- and $b$-axis for all compositions look very similar showing mainly a paramagnetic Curie-Weiss behavior of Tb-moments down to \TNTb~as shown in the inserts in Fig.~\ref{MvsT-data}. The anomaly corresponding to onset of Tb-magnetic ordering at \TNTb~shifts to lower temperatures and becomes broader with increasing Ga content. 

\begin{figure}[bt!]
\begin{center}
\includegraphics[width=85mm]{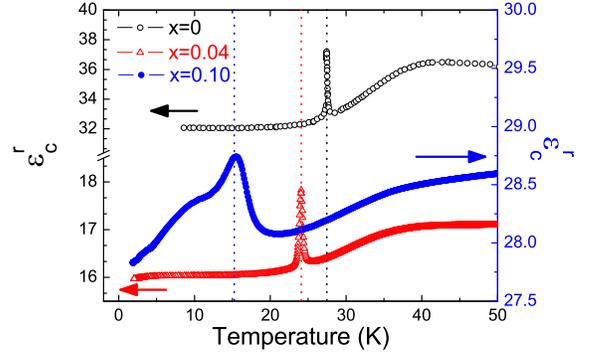}
\caption{Temperature evolution of relative dielectric constant  ($\epsilon^{r}_{c}$)  for TbMn$_{1-x}$Ga$_{x}$O$_{3}$ with $x$ = 0, 0.04, 0.1, respectively, measured on cooling with $E=20$ kHz along the $c$-direction. The data for $x=0$ are extracted from Ref.\cite{kimura3} }
\label{fig5}
\end{center}
\end{figure}

Having observed a rather similar magnetic behavior of the Mn-sublattice in the pure and the substituted TbMnO$_{3}$ compounds, we have performed dielectric measurements for the $x = $ 4\% and 10\% samples. Impedance spectroscopy measurements from 1 kHz up to 20 kHz reveal no relaxor behavior for all compounds. Similar to TbMnO$_{3}$, the 4\% and 10\% Ga samples present a strong anisotropy and anomaly in the dielectric constant along $c$- and $a$-directions at 40~K. In TbMnO$_{3}$ the ferroelectric transition is characterized by a lamda-shaped anomaly in the $c$-direction at 27~K (Fig.~\ref{fig5}).\cite{kimura3} The dielectric measurements along the $c$-direction for 4\% and 10\% samples show similar anomalies shifted to lower temperatures. They appear at 24~K and 15~K, respectively, in perfect agreement with $T_{\rm FE}$ determined from magnetization measurements. The height of the lamda-shaped anomaly of the dielectric constant ($\delta\epsilon$) is reduced with increasing Ga content from $\delta\epsilon$ of about 4 for $x=$ 0 down to 1.8 and 1 for $x=$4\% and 10\%, respectively. All these features suggest that at the respective $T_{\rm FE}$ the compounds undergo a ferroelectric transition that becomes less pronounced with increasing Ga-doping.

The results from the measurements of magnetization and dielectric constant as function of temperature are summarized in Fig.~\ref{fig1}c. It shows that substitution of nonmagnetic Ga$^{3+}$ for magnetic Mn$^{3+}$ leads to reduction of $T_{\rm N}^{\rm Mn}$ linearly with a slope of -0.6~K per 1$\%$ of Ga. This is direct evidence of the effective reduction of the strength of Mn-Mn interactions, $J_{\rm Mn-Mn}$ and, consequently, we suggest that $J_{\rm Mn-Tb}$ should be reduced. $T_{\rm FE}$ gets also reduced by -1.2 K per 1$\%$ of Ga.
The most surprising effect, however, is that the Ga for Mn substitution affects the anomalies at \TNTb. Indeed, the larger $x$ is the smaller and broader becomes the anomaly corresponding to the transition to Tb-magnetic ordering (see Fig.~\ref{MvsT-data}). 

\begin{figure}[bt!]
\begin{center}
\includegraphics[width=98mm]{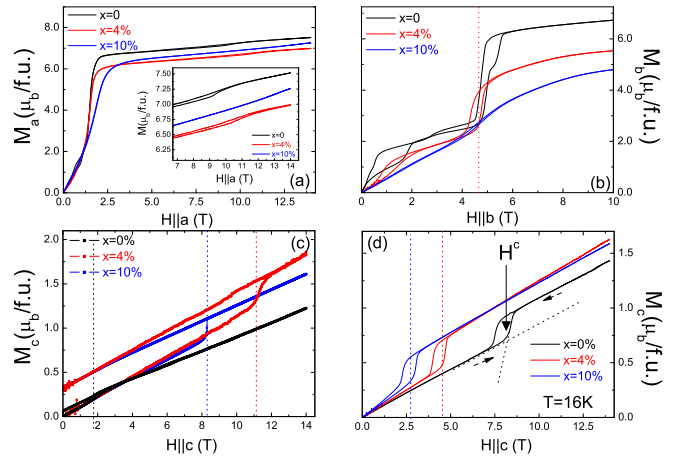}
\caption{Magnetization isotherms of TbMn$_{1-x}$Ga$_{x}$O$_{3}$ ($x=$ 0, 0.04 and 0.1) single crystals with magnetic field along $a$- (a), $b$- (b) and $c$-crystallographic (c) directions at $T=2$ K. The insert in the panel a) shows a zoomed part of \Ha\space magnetization curve, respectively. The panel d) presents the magnetization isotherm at $T=16$ K for \Hc. After each hysteresis loop the samples have been heated up to 160 K and then zero field cooled down to the target temperature.}
\label{MvsH}
\end{center}
\end{figure}

Further we studied the magnetic field response of the Ga-substituted compounds. Fig.~\ref{MvsH} shows isothermal magnetization curves as a function of magnetic field for all samples. The data on undoped TbMnO$_{3}$ compound is in good agreement with previously published results.\cite{kimura3} For $H\|a$ at 2 K (Fig.~\ref{MvsH}a), two magnetic transitions are observed at $H_{\rm c}\sim$ 1.7 T and 9 T. The critical fields of all meta-magnetic transitions, $H_{\rm c}$, have been defined by the intersection point between two lines: the first being a linear fit of $M(T)$ before the anomaly; the second passing through the inflection point of the anomaly as shown, for example, in (Fig.~\ref{MvsH}d). The low field metamagnetic transition is a transition to the field forced ferromagnetic state associated with the Tb-sublattice. Similar meta-magnetic rare earth behavior is reported for the isostructural TbAlO$_3$ compound.\cite{holmes} One marks out that the magnetization is not saturated up to 14~T and the extracted value of the Tb-moment, $\sim$ 6.5~$\mu_{\rm B}$, is rather low.  All these suggest that the Tb moments do not become fully ferromagnetically ordered above 2~T. Instead, because of their strong Ising anisotropy, they presumably arrange in a noncollinear (F$_x$~C$_y$~0) pattern (Bertaut notation)\cite{Bertaut} with $m_x \sim$ 6.5~$\mu_{\rm B}$. Assuming theoretical 9~$\mu_{\rm B}$/Tb$^{3+}$, that would correspond to 36$^\circ$ canting to $a$-axis in good agreement with reported data on Tb-anisotropy.\cite{Quezel,bielen} Much less pronounced second transition corresponds to the flop of the electric polarization from \pc\space to \pa~that is presumably caused by the flop of Mn cycloidal plane.\cite{kimura3,nadir2}

Both meta-magnetic transitions are observed in $x=4$\% and $x=10$\% samples.
The field forced Tb-ferromagnetic ordering is observed at slightly changed critical fields of 1.8 T and 2.4 T for for $x=4$\% and $x=10$\%, respectively, while the second (polarization flop) transition is found at higher field $H_{\rm c}=11$ T for $x=4$\% (see Fig.~\ref{MvsH}(a) insert). A change in slope is also observed for $x=10$\% around 14 T suggesting that the second transition might occur at even higher fields.

For $H\|b$\space at 2 K, also two meta-magnetic transitions are observed in TbMnO$_{3}$ in good agreement with  published data.\cite{kimura3} The first one at 1.7 T is a transition to a magnetic state where Tb orders with $\bf{\tau}=$ 1/3 \bb.\cite{nadir} The second transition is observed at around $H\|b \sim$ 4.9 T. Based on the data on TbAlO$_3$,\cite{holmes} we suppose that Tb-spins undergo a transition to noncollinear (C$_x$~F$_y$~0)-arrangement in this state. According to Ref.\cite{nadir2}, this transition is associated with the flop of Mn-cycloidal plane from $bc$ to $ab$ causing the electric polarization flop from \pc\space to \pa.\cite{kimura3,nadir2} Both meta-magnetic transitions are observed in $x=4$\% sample. They are shifted to lower critical fields of 1.0 T and 4.6 T, respectively. Only one clearly separated transition with $H_{\rm c}=4.1$ T is observed for $x=$ 10\% sample.

For $H\|c$, TbMnO$_{3}$ shows one metamagnetic transition as visualized in Fig.~\ref{MvsH}(c-d) for 2 and 16 K, respectively. It corresponds to the transition from cycloidal Mn-magnetic structure to simple canted antiferromagnetic structure and, simultaneously, to the transition from ferroelectric to paraelectric state.\cite{ArgyriouHc,kimura3} Both substituted compounds show similar metamagnetic transitions at lower critical fields, Fig.~\ref{MvsH}(c-d).

In short summary, bulk characterization of the Ga-substituted compounds revealed rather similar magneto-electric phenomena in all the studied samples. The only crucial differences are applied to the Tb-magnetic ordering, where strong suppression of relevant anomalies has been observed. To clarify what causes these changes, a microscopic view on Mn- and Tb-magnetic orders in the Ga-samples has been taken.

\subsection{Magnetic ordering}

\begin{figure}[bt!]
\begin{center}
\includegraphics[width=85mm]{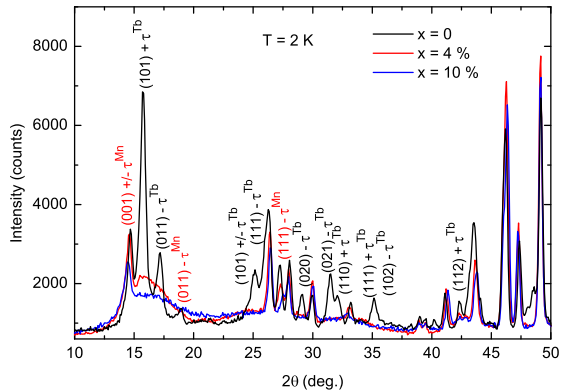}
\caption{(Color online) Neutron diffraction patterns of TbMn$_{1-x}$Ga$_{x}$O$_{3}$ compounds as measured with E9 diffractometer ($\lambda = $2.816~\AA) at 2 K.}
\label{fig3}
\end{center}
\end{figure}

\begin{figure}[bt!]
\begin{center}
\includegraphics[width=85mm]{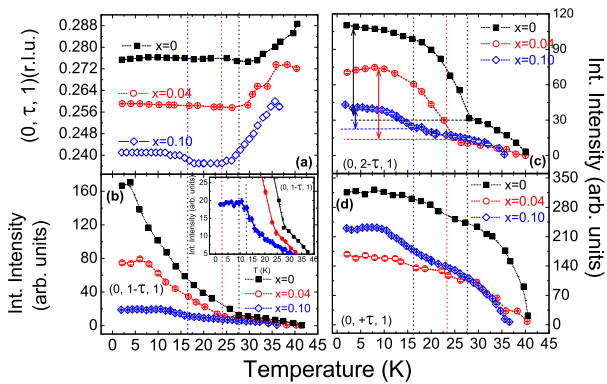}
\caption{ a) Temperature evolution of the propagation vector $\tau^{\rm Mn}$ of
TbMn$_{1-x}$Ga$_{x}$O$_{3}$ with $x$ = 0, 0.04, 0.1 collected with the E1 triple axis spectrometer. Panels b-d) represent the temperature evolution of the integrated intensity of G-type (0 1-$\tau$ 1), A-type (0 $\tau$ 1) and A-type (0 2-$\tau$ 1) incommensurate magnetic reflections, respectively. The insert in the panel b) zooms the temperature evolution of integrated intensity of the G-type reflection around the $T_{\rm FE}$. The data have been collected on heating.}
\label{tauvsT}
\end{center}
\end{figure}

Microscopic insight into the magnetic ordering in TbMn$_{1-x}$Ga$_{x}$O$_{3}$ was obtained using ND on powder (Fig.~\ref{fig3}) and single crystal samples (Fig.~\ref{tauvsT}). XRMS has been used to contrast Tb-magnetic signal when it contributed to the same reflections as Mn. Cooling the samples below $T_{\rm N}^{\rm Mn}$ we found a set of magnetic satellites in Brillouin zones ($hkl$) with extinction conditions $h+k=even$, $l=odd$, known as A-type reflections.\cite{Bertaut} They arise from the ordering of Mn-spins that is incommensurate (ICM) and characterized by the propagation vector $\mbox{\boldmath$\tau$}^{\rm Mn}$ $=(0~\tau^{\rm Mn}_y~\space0)$. The temperature dependence of the propagation vector is shown in Fig.~\ref{tauvsT}a. It decreases linearly with decreasing temperature from $T_{\rm N}^{\rm Mn}$ down to $T_{\rm FE}$ for all compositions. Below $T_{\rm FE}$ $\tau$ stays constant. 
With increasing Ga-content $\tau^{\rm Mn}$ decreases (Fig.~\ref{fig1}c) that is directly related to change of the  Mn-O-Mn bond angle. One also observes the reduction of the overall magnetic satellite intensities as expected from the reduced total magnetization. 

The A-type satellites are typical for all studied compositions in temperature range from $T_{\rm N}^{\rm Mn}$ down to 2~K, showing that the  main details of the Mn magnetic structures for all compounds stay similar. 
Using the representation analysis,\cite{Bertaut} the Mn antiferromagnetic ordering below $T_{\rm N}^{\rm Mn}$ can be described by one single component of the irreducible representation $\Gamma_{3}(0, A_{y},0)$. This gives rise to an amplitude modulated AF Mn spin ordering along the $b$-direction below $T_{\rm N}^{\rm Mn}$.\cite{Quezel}

Below $T_{\rm FE}$ the Mn-magnetic structure of TbMnO$_3$ has been described by the linear combination of two irreducible representation $\Gamma_{3}\oplus\Gamma_{2}$ where $\Gamma_{2}$ has also A-component along $c$-axis $(0, 0, A_{z})$.\cite{kenz} The resulting structure is a characteristic for ferroelectric phase: cycloid $m$(Mn)=$(0,m_{y},m_{z})$, where Mn-spins rotate within $bc$-plane.  
The fact that the coupling between Fourier components $m_z$ is also of A-type complicates the structure analyses as both $m_y$ and $m_z$ contribute to the same A-type reflections. However, following a temperature dependence of a single magnetic reflection often helps to see an anomaly at temperature ($T_{\rm FE}$ in our case) when the second magnetic component develops. Indeed, the temperature evolution of the integrated intensity of the selected (0, 2-$\tau^{\rm Mn}$, 1) and (0, $\tau^{\rm Mn}$, 1) plotted in Fig.~\ref{tauvsT}c,d shows an inflection point at 27.5 K for $x=0$ compound in a good agreement with $T_{\rm FE}$. 
Similar considerations applied to the Ga-doped samples allow to conclude that their Mn-structures have a second magnetic component of A-type developing below
25~K and 17~K for $x=$0.04 and 0.1 Ga-doped samples, respectively. Both temperatures are again in perfect agreement with $T_{\rm FE}$ determined from the bulk measurements. The arrows in Fig.~\ref{tauvsT}(c) show that the relative amount of the spin along the $c$-direction decreases with increasing Ga-content.
The results of magnetic structure powder refinement assuming cycloidal magnetic ordering $m$(Mn)=$(0,m_{y},m_{z})$ are presented in Table~\ref{table1}.  

Coupling between Tb- and Mn-magnetic orders manifests itself below $T_{\rm N}^{\rm Mn}$ where Mn-spin lattice polarizes the rare magnetic sublattice.\cite{kenz,nadir-review} This is reflected in appearance of G-, C- and F-type reflections that become especially strong below $T_{\rm FE}$. Fig.~\ref{tauvsT}b shows the temperature dependence of selected G-type reflection for all compositions. Unlike the A-type reflections, this
one has a pronounced concave shape as a characteristic of induced order. With increasing Ga-content the intensity of induced reflections drastically decreases. Since intensity is proportional to the squared magnetic moment, ${\bf M}_{\perp}({\bf Q})^2$, one can conclude that the effective field acting on the Tb-moments from the Mn-spin structure is indeed reduced upon Ga-substitution.

\begin{figure}[bt!]
\begin{center}
\includegraphics[width=88mm]{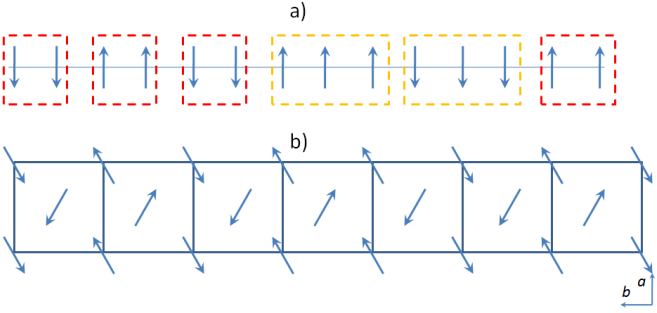}
\caption{Magnetic unit cell of Tb-sublattice in the $ab$-plane (in the adjacent planes along $c$-axis the magnetic moments are antiparallel): a) The structure proposed by one-dimensional ANNNI Ising model from Ref.\cite{prok2};  b) Magnetic structure model taking into account Tb-anisotropy}
\label{structure}
\end{center}
\end{figure}

Below $T_{\rm N}^{\rm Tb}=$ 7 K, undoped TbMnO$_3$ shows very intense magnetic satellites appearing at positions characterized by a propagation vector $\tau^{\rm Tb}$= 0.426(2) \bb ($\sim 3/7$) and its odd harmonics (Fig.~\ref{fig3}). Quezel $et~al.$\cite{Quezel} proposed a sine-wave structure with the Tb-moments lying along two symmetrical directions with respect to the $b$-axis (57$^\circ$) within the $ab$-plane. According to the results from our previous work,\cite{prok2} in one-dimensional case the Tb-magnetic structure can be understood as being built from $\uparrow \uparrow \downarrow \downarrow$-blocks in the $ab$-plane 
with some irregularities to adjust Ising-like Tb spins to periodic Mn ordering (Fig.~\ref{structure}a). Taking into account data on rare earth behavior in isostructural 3$d$-less aluminates $R$AlO$_3$\cite{bielen,bouree}~and similar manganites with $R=$ Dy and Ho,\cite{prok,munoz} that all show the same type of G$_x$A$_y$ ordering with $\tau=0$ and $\tau=1/2$, respectively, one can make this model more realistic. Indeed, taking G$_x$A$_y$-coupling to preserve Tb-anisotropy 
and keeping the same periodicity as in Fig.~\ref{structure}a, one gets a structure visualized in Fig.~\ref{structure}b. This structural model has been used to fit our powder diffraction data and the results of the refinement are presented in Table~\ref{table1}. One has to stress, however, that this solution is not unique and the real structure might be even more complex. On the other hand, it accounts for both the observed harmonic clamping of Tb- and Mn-propagation vectors at low temperatures\cite{prok2} and the Tb-anisotropy.\cite{bielen}

One of the most striking effects of Ga for Mn substitution appears below $T_{\rm N}^{\rm Tb}$ where magnetic ordering of the Tb-sublattice is expected (Fig.~\ref{fig3}). Even the powder ND pattern of the 4$\%$ substituted compound looses all the features of long range Tb magnetic ordering: Tb superlattice reflections are replaced by low intense and broad peaks visible predominantly at low angle positions of the strongest Tb-peaks. With increasing Ga-content the absolute intensity of Tb peaks drops and their width increases (Fig.~\ref{fig1}d). Altogether, the observed behavior reflects the destruction of long range Tb-magnetic ordering and its substitution by short range magnetic clusters with correlation length less than 25~\AA\ (Fig.~\ref{fig1}d).

The same effect is found by XRMS at the L$_3$ absorption edge of Tb. In particular, XRMS data on TbMn$_{1-x}$Ga$_{x}$O$_{3}$ single crystals showed the presence of induced Tb magnetic ordering with the Mn propagation vector $\tau^{\rm Mn}$=0.262 \bb~at $2 < T <28$~K for $x=4$\%. The observed magnetic reflections are about 30 times weaker that those of pure TbMnO$_3$ and exist down to the lowest temperatures where no additional reflections corresponding to Tb long range magnetic ordering (i.e. with $\mathbf{\tau^{Tb}}$) have been observed. For $x=10$\%, XRMS found no Tb-magnetic reflections in the full temperature range below $T_{\rm FE}$. Single crystal ND, however, reveals some anomalies in temperature dependence of the Mn propagation vector and integrated intensity and width of ICM reflections. At 17~K on heating, $\tau$ suddenly decreases (Fig.~\ref{tauvsT}a) and, simultaneously, the width of the all the Mn ICM reflections gets smaller indicating a sudden change of the Mn-Mn vs Mn-Tb magnetic correlation. 

\section{ Discussion}

\begin{figure}[bt!]
\begin{center}
\includegraphics[width=75mm]{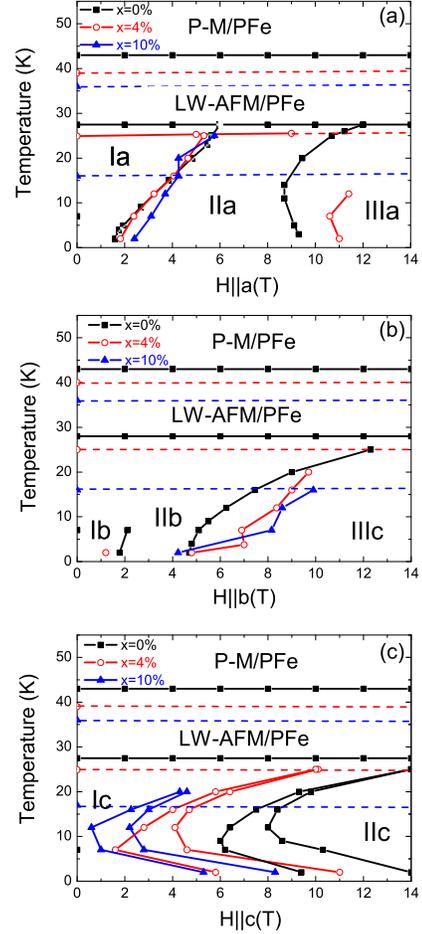}
\caption{Magnetic phase diagram of TbMn$_{1-x}$Ga$_{x}$O$_{3}$ compounds $x$ = 0, 0.04, 0.1 and magnetic fields applied along the \Ha~(a), \Hb~(b) and \Hc~(c) axes. The squares, circles and triangles represent the data obtained by measurements of magnetization and dielectric constant for $x=0$, 4 and 10\% samples, respectively. P-M/PFe defines paramagnetic and paraelectric phase,  while LW-AFM/PE corresponds to the phase where the Mn has an antiferromagnetic long wave ordering and still paraelectric state. The phases Ia, Ib and Ic are identical and correspond to $bc$-cycloidal Mn-ordering accompanied by $P_{\rm S}$ along the $c$-direction. IIa is field forced ferromagnetic ordering  of Tb while IIb is an antiferromagnetic ordering of Tb with $\tau=\frac{1}{3}$. The IIIa and IIIb phases define $ab$-cycloidal Mn-ordering with \Ps~along $a$-direction (for $x=0$, at least). The IIc zone is paraelectric phase with simple Mn-antiferromagnetic ordering. The dashed lines are expected phase boundaries for the Ga-containing compounds. The dielectric phase boundary for $x=0$ has been extracted from Ref.\cite{kimura3}}
\label{Diagram}
\end{center}
\end{figure}

Based on the results from magnetic and dielectric, neutron and x-ray diffraction measurements, a magnetic phase diagram of TbMn$_{1-x}$Ga$_{x}$O$_{3}$ compounds with $x$ = 0, 0.04, 0.1 has been created (Fig.~\ref{Diagram}). For small Ga doping there is no dramatic change in the magnetic order of the system.\footnote{Minor changes in Mn-magnetic order allow to neglect another unavoidable effect of Ga for Mn substitution: disturbance of Mn orbital order.} 
From the dielectric measurements along the $c$-axis we see that Ga doping also does not change the ferroelectric state. However, substitution provides a platform for studying the interplay between the Mn- and Tb-magnetic states.

According to the phase diagram in Fig.~\ref{Diagram}, all the magnetic phases persist for $x=4$\% and 10\% compounds. The main differences are found in the magnetic state of Tb and critical temperatures and fields of the corresponding phase transitions. Let us start with Tb-magnetic ordering and corresponding effects of Ga for Mn substitution. Our results unambiguously show that $J_{\rm Mn-Tb}$ is important ingredient for Tb-magnetic order both below and above $T_{\rm N}^{\rm Tb}$.
The induced Tb-magnetic ordering ($\tau^{\rm Tb} = \tau^{\rm Mn}$ above $T_{\rm N}^{\rm Tb}$), should be strongly affected by Ga for Mn substitution reducing $J_{\rm Mn-Tb}$. This finds a direct confirmation in our data as both $T_{\rm FE}$ and the intensity of the induced Tb-magnetic peaks (i.e. averaged induced moment) are significantly reduced upon Ga substitution.

Below $T<T_{\rm N}^{\rm Tb}$, when $J_{\rm Tb-Tb}$ competes with $J_{\rm Mn-Tb}$, the system minimizes its energy through the matching of Tb- and Mn-wave vectors, $3 \tau^{\rm Tb} - \tau^{\rm Mn} = 1$.\cite{prok2} The resulting Tb-magnetic structure is presented in Fig.~\ref{structure}. Clearly, this magnetic regime is supposed to be also strongly affected by the Ga for Mn substitution. Our data strongly support this picture showing that the substitution of magnetic by nonmagnetic ions in one of the sublattices can change the magnetic ordering in the other sublattice. Here, Tb-magnetic ordering becomes of short range in a very narrow 0 $< x\leq $ 0.04 range of Ga for Mn substitution. First of all, such low critical concentration explains scatter in the published data on Tb-ordering\cite{Quezel,kajimoto,kenz} since small non-stoichiometry in the samples would lead to a significant effect on it. Secondly, disappearance of Tb long range magnetic ordering proves that Mn exchange fields are involved in Tb magnetic ordering in TbMnO$_3$ even below $T_{\rm N}^{\rm Tb}$.

Ideally, one can expect by reducing $J_{\rm Mn-Tb}$ to get an independent Tb-magnetic ordering determined only by $J_{\rm Tb-Tb}$. Since TbGaO$_3$ does not exist, one can again look at 3$d$-less Tb- and DyAlO$_3$ both showing the same type of G$_x$A$_y$ ordering in $ab$-plane.\cite{bielen,bouree} Mn-containing DyMno$_3$ and HoMnO$_3$ show the same type of magnetic ordering but with doubling of unit cell along $b$-direction.\cite{prok,munoz} Expanding the same considerations on TMnO$_3$, one can expect that strong $J_{\rm Mn-Tb}$ prevents Tb of having regular $\tau^{\rm Tb}$=1/2 \bb~structure presumably of G$_x$A$_y$-type (in one-dimensional case it would simply correspond to $\uparrow \uparrow \downarrow \downarrow$). 

This, however, does not happen since no long range Tb-order has been observed at low temperatures. 
We believe the reason for that is Ga for Mn substitution does not correspond to homogeneous reduction of Mn-Tb exchange but rather to pronounced effects located at the sites which are occupied by Ga.
In this case one has to distinguish between Tb ions in the nearest neighborhood from Ga and not. The former get released from Mn-exchange fields, as seen by the significantly reduced average induced moment below $T_{\rm FE}$, and  can form their own $\tau^{\rm Tb}$=1/2 \bb~ordering below 7 K. However, the latter follow the Mn periodicity above 7 K and, consequently, keep '$3 \tau^{\rm Tb} - \tau^{\rm Mn} = 1$'-state below 7 K (Fig.~\ref{structure}b). As a result one can still expect a kind of G$_x$A$_y$ ($\uparrow \uparrow \downarrow \downarrow$) preserved at a local scale while the overall periodicity is completely broken. 

This model finds a strong support in the low field data on doped TbMn$_{1-x}$Ga$_{x}$O$_{3}$ with \Ha\space and \Hb\space (Fig.~\ref{Diagram}). Indeed, these data show that there are no principal differences to the undoped compound. With \Ha\space and \Hb, the low field phase boundary stays almost unaffected (except of \Hb~for $x=10$\% sample where the transition becomes very smooth). Having in mind, that this transition in TbMnO$_{3}$ corresponds to magnetic field forced noncollinear F$_x$C$_y$ or C$_x$F$_y$ arrangement for \Ha~and \Hb, correspondingly, one finds it consistent with Tb-ordering of G$_x$A$_y$ ($\uparrow \uparrow \downarrow \downarrow$) preserved at local scale in Ga-compounds. The high field boundaries for \Ha, \Hb, both getting increased upon Ga substitution, is more difficult to interpret. Intuitively, one would expect the reduction of both critical fields. However, to answer this question one has to know the magnetic and electric states of these compounds above these critical fields. This issue stays outside the topic of present publication.

Finally, we summarize the effects of Ga for Mn substitution on Mn-sublattice. Since it is the Mn-sublattice that has been diluted, one can straightforward interpret the changes directly involving Mn-magnetism. At first, one marks out here  the reduction of $T_{\rm N}^{\rm Mn}$ and total magnetization reflecting the reduced $J_{\rm Mn-Mn}$. 
Neutron diffraction confirms that the magnetic structures observed below these temperatures are similar while dielectric measurements show the anomalies pointing to ferroelectric character of transitions at $T_{\rm FE}$ for all samples. Among magnetic isotherms one can have a look at the \Hc~data, as Tb-contribution in there is minimal. The critical field $H_{\rm C}$ along $c$-axis in TbMnO$_{3}$ corresponds to the phase transition from ICM Mn-cycloid to a simple commensurate antiferromagnetic ordering. Substituting nonmagnetic Ga for Mn should result in the reduction of $H_{\rm C}$ as has been observed in the experiment (see Fig.~\ref{MvsH}). Generally, one can conclude that all the changes in the magnetic properties of Mn-sublattice are mainly of quantitative character and in agreement with effective reduction of $J_{\rm Mn-Mn}$.

\section{ Conclusions}

Ga for Mn substitution in multiferroic TbMnO$_{3}$  has been performed in order to investigate experimentally the influence of Mn-magnetic ordering on the Tb-magnetic sublattice. Single and polycrystalline TbMn$_{1-x}$Ga$_{x}$O$_{3}$ ($x$ = 0, 0.04, 0.1) compounds have been synthesized and characterized by powder and single crystal neutron diffraction, x-ray resonant magnetic scattering, single crystal magnetization and dielectric measurements. The results show that light Ga for Mn substitution does not change qualitatively the magnetoelectric properties of the   Mn-sublattice while it significantly affects the Tb-magnetic ordering. The latter looses its long range character in a small range of Ga-concentration while preserves the magnetic structure locally. Thus, our work proves that there is a strong influence of Mn-magnetism on Tb-magnetic ordering kept down to the lowest temperatures. The whole ordering scheme of Tb in multiferroic TbMnO$_{3}$ results from competing  $J_{\rm Mn-Tb}$  and $J_{\rm Tb-Tb}$ exchange interactions verifying our theoretical model reported in Ref.\cite{prok2}

\section{ Acknowledgements}

The authors have benefited from discussions with D. Khomskii, M. Mostovoy and K. Prokes. This work was supported by the Hytrain project of the Marie Curie research Training Network funded under the EC's 6th Framework Human Resources. D.N.A. thanks the Deutsche Forschungsgemeinschaft for financial support under Contract
No. AR 613/1-1.

\end{document}